# One Dimensional Nearly Free Electron States in Borophene

*Longjuan Kong[1,2#], Liren Liu[3#], Lan Chen[1,2,4*], Qing Zhong[1], Peng Cheng[1,2], Hui Li[5], Zhuhua Zhang[3*], and Kehui Wu[1,2,4*]*

[1]Institute of Physics, Chinese Academy of Sciences, Beijing 100190, China

[2]School of physics, University of Chinese Academy of Sciences, Beijing 100049, China

[3]State Key Laboratory of Mechanics and Control of Mechanical Structures, and Key Laboratory for Intelligent Nano Materials and Devices of Ministry of Education, Nanjing University of Aeronautics and Astronautics, Nanjing 210016, China

[4]Songshan Lake Materials Laboratory, Dongguan, Guangdong 523808, China

[5]Beijing Advanced Innovation Center for Soft Matter Science and Engineering, Beijing University of Chemical Technology, Beijing 100029, China.

[#] These authors contributed equally to this work

**ABSTRACT** Two-dimensional boron (borophene) is featured by its structural polymorphs and distinct in-plane anisotropy, opening opportunities to achieve tailored electronic properties by intermixing different phases. Here, using scanning tunneling spectroscopy combined with first-principles calculations, delocalized one-dimensional nearly free electron states (NFE) in the (2,3) or $\beta_{12}$ borophene sheet on the Ag(111) surface were observed. The NFE states emerge from a line defect in the borophene, manifested as a structural unit of the (2,2) or $\chi_3$ sheet, which creates an in-plane potential well that shifts the states toward the Fermi level. The NFE states are held in the 2D plane of borophene, rather than in the vacuum region as observed in other nanostructures. Furthermore the borophene can provide a rare prototype to further study novel NFE behaviors, which may have potential applications on transport or field emission nanodevices based on boron.

Dispersive nearly free electron (NFE) states on material surfaces have drawn persistent interest in the past years, due to the remarkable electron transport properties and potential applications in electron emitters.[1-7] NFE states are usually unoccupied states, with typically parabolic energy dispersion characterized by an effective mass nearly identical to the free electron mass ($m_e$),[8-10] and widely exist on low-dimensional materials due to the confinement potential normal to the surface, such as two-dimensional (2D) graphite,[11, 12] graphene,[13] transition-metal dichalcogenides,[14] one-dimensional (1D) nanoribbons and nanotubes,[15, 16] and even zero-dimensional $C_{60}$ molecules.[3, 17] Moreover, these NFE states are usually extended in the vacuum region above the material surfaces rather than reside at the basal plane, which however limits their influences on the transport properties. To our best knowledge, the NFE states inside a material have not been reported.

Recently, a new 2D material from group-III element boron, named as borophene, has been experimentally realized. Borophene possesses intriguing electronic and mechanical properties that are attractive for potential applications in nanodevices.[18-30] Due to the electronic deficiency, boron can form both classical covalent bonds and a variety of multicenter bonds. The multicenter bonds not only compensate the electronic deficiency of boron, but also result in a structural diversity of borophene,[31, 32, 33, 34, 35] which was predicted by density functional theory (DFT) calculations,[36] and then verified by experimental observations of different phases on Ag(111),[18, 19, 37] Ag(110),[38] Al(111),[22] and recently on Cu(111) surfaces,[39] indicating highly similar thermal stability of these phases. It is interesting that except the honeycomb borophene phase on Al(111), all other borophene phases reported so far adopt striped structures, whose building blocks are ribbons of triangular lattices separated by single rows of hollow hexagons (HH). Based on various widths of ribbons, these striped borophene structures can be indexed as (m, n), where m, n denote the number of atoms in the narrowest and widest regions of a single boron ribbon, respectively (shown in Figure 1f).[38] The coexistence of these borophene phases suggests that one can construct in-plane homojunctions of borophene by intermixing ribbons with different widths, which was also experimentally observed recently.[40] Such homojunction structures combined with the metallicity of borophene may lead to exotic electronic behaviors.

Here we report on the investigations of a borophene homojunction on Ag(111), which is comprised of two (2,3) domains (i.e. the $v_{1/6}$[34] or $\beta_{12}$[18] sheet) separated by two (2,2) ribbons as line defect. Scanning tunneling microscopy/spectroscopy (STM/STS) characterizations reveal pronounced and delocalized electronic states above the Fermi level near the line defect, which are well reproduced by first-principles calculations. The energy levels, parabolic energy dispersion, and extended spatial distribution unambiguously support the NFE nature of these states, attributed to an in-plane potential well within the boron monolayer induced by the line defect. The 1D NFE states, spatially held to the plane of the borophene sheet instead of in the vacuum region and further amenable to modulation by applied in-plane strain, may be of great relevance to applications of borophene devices.

**Experiments and Methods.** Borophenes were grown on the Ag(111) surface in an ultra-high vacuum (UHV) chamber (p ≈ 5.0×10$^{-11}$ torr) by direct evaporation of a pure boron source. The single crystal Ag(111) surface was cleaned with several Ar$^+$ sputtering and annealing cycles. The substrate was preheated to about 570 K before boron deposition. After growth, the sample was *in-situ* transferred to a combined chamber for STM/STS characterizations. All STM/STS measurements were performed at liquid helium temperature (~ 5 K) by a chemically etched tungsten tip. The differential conductance (d$I$/dV) spectra were measured by superimposing an ac voltage (10 mV, 669 Hz) on the given dc bias, and recording the in-phase ac component in the tunneling current with a lock-in amplifier. The bias voltage was applied to the tip.

The first principles calculations were performed within the framework of density functional theory with generalized gradient approximation (GGA) of the Perdew-Burke-Ernzerhof (PBE) functional,[41] as implemented in the Vienna Ab initio Simulation Package (VASP).[42,43] Firstly, the structural relaxation is applied on intermixing borophene superlattice with 19 ribbons of (2,3) phase and 2 ribbons of (2,3) pure phase boron on the Ag(111) substrate with three-layer thickness, as shown in Figure S1a) in the Supporting Information. The fully optimized lattice parameters are $a_0$ = 105.08 Å, $b_0$ = 2.89 Å, with periodic boundary condition at these two directions. The side view of borophene film is flat, indicating a weak hybridization between borophene and Ag(111) surface, and thus representing a quasi-freestanding sheet. Hence, in order to reduce the computational cost, the Ag(111) surface is omitted in the DFT calculation of electrical properties. Unless specified otherwise, the corresponding properties we referred to are calculated by freestanding borophene with line defect (in Figure S1b)). A vacuum space of 20 Å was set to eliminate spurious interaction between two borophene sheet in adjacent periodic images. The core electron-ion interaction was described by the projector augmented wave (PAW) potential,[44] and the plane-wave basis set cutoff energy was set to be 450 eV. The Brillouin zone integration was sampled using 2×4×1 $k$ points for the structural optimization and 20 $k$ points for band structure calculations. The atomic positions were optimized until the maximum force on each atom was less than 0.01 eV/Å.

**Results and Discussions.** Borophene grown on Ag(111) with substrate temperature around 570 K usually exhibits the (2,3) phase, i.e. the $v_{1/6}$ or $\beta_{12}$ sheet.[18] Figure 1a is a STM image showing one borophene island of (2,3) phase on Ag(111) surface, with three domains separated by linear boundaries (denoted by the white arrows). The high-resolution STM image, as shown in Figure 1e, clearly reveals the atomic structure of the boundary, where one dark stripe separates two borophene domains. The two domains appear mirror-symmetric with the dark stripe as the mirror plane. We also found that the dark stripe has the same lattice constant as that of borophene domains, but relatively shifted half unit cell along the direction of stripes. The two stripes neighboring to the dark stripe appear brighter, while their lattice position remains consistent with the (2,3) borophene domains.

Note that the contrast of the boundary region in STM image highly depends on the bias, indicating that there are specific electronic states around the boundary. We performed STS measurements around the boundary. Figure 1b-d are the d$I$/dV maps of the same borophene island as in Figure 1a, which reveal the drastic change of the local density of states (LDOS) with tip bias around the domain boundary. Importantly, when the tip bias is small (< 3 eV), the boundary state appears localized at the boundary (Figure 1c,d). In contrast, at high energy range (> 3 eV) a pronounced delocalized DOS is observed, spreading to several ribbons to the interior of domain, as show in Figure 1b. To quantitatively probe the spatial distribution of this state, we measured the d*I*/dV spectra on spots of (2,3) phase with various distances from the line defect, which are labeled by the dot line as shown in the upper part of Figure 2a. Two dominant peaks in the d*I*/dV curves shown in Figure 2a appear at the bias range from -2 to -4 V, corresponding to the energy region 2 ~ 4 eV above the Fermi level. The peak at -2.5 V has position-independent intensity, corresponding to an intrinsic electronic band of the borophene (2,3) phase, which has also been reported before.[18] On the contrary, the peak near -3.4 V exhibits a nonlinear decaying behavior with increasing distance from the line defect. Specifically, the peak is absent at the brighter stripe nearest to the line defect (No.0 stripe shown in upper part of Figure 2a). It starts to develop at the first (2,3) boron ribbon (No.1), and the density increases and reaches the maximum at the fourth (2,3) ribbon (No.4). Then it slowly decays and disappears at the 10th (2,3) ribbon (No.10), which is deep into the interior of the (2,3) borophene domain. Meanwhile, the energy position of this peak also slightly shift to low energy when moving away from the boundary. The delocalized spatial distribution of the LDOS around 3.4 eV above Fermi level is consistent with d*I*/dV maps (Figure 1b). The lateral spatial distribution of this electronic state indicates that it is neither an intrinsic state of pure borophene sheet of (2,3) phase, nor due to the interactions with Ag(111) substrate, since in those cases the state should be laterally uniform.

To understand the experimental results, we performed first-principles calculations to identify the atomic model and elucidate the origination of the delocalized electronic states. First, after an extensive search for structural models, we found that the features around the line defect are well captured by two interconnected (2,2) ribbons (a structural unit of the $v_{1/5}$ sheet), which also share the same lattice periodicity as the (2,3) phase along the HH rows. The atomic model (depicted in Figure 1g is further supported by the overall agreement between the simulated STM image of the (2,3) phase interconnecting two (2,2) ribbons and the experimental image (Figure 1e,f). It should be noted that the Ag substrate is included for STM simulation and a very huge unit cell (containing 21 boron stripes) is constructed to look for the precisely spatial distribution of LDOS. However, we believe the weak interfacial interactions can not alter the band structures of borophene strongly, while the charge transfer from Ag surface to borophene will cause the shift of Fermi level.

Based on the aforementioned structural model (two (2,3) domains interconnected by two (2,2) ribbons), we calculated the electron density of states projected to boron dimers on different boron stripes, marked by number 0-10 along a line perpendicular to the line defect, as shown in Figure 2b. Several peaks appear in the energy range of 1.5-3.0 eV. The major peaks located at 2.0-2.5 eV are attributed to normal π* electronic states and therefore do not exhibit clear spatial dependence. In contrast, we also identify a peak at energy of ∼3.9 eV in the PDOS corresponding to the No.1 stripe. This state is found to exhibit a strong spatial dependence and is blurred when moving away from the line defect. These features agree well with the experimentally observed variations of both the energy location and intensity, except that the experimentally observed state is located at ∼ -3.4 V in the d$I$/dV spectra, slightly lower than the calculated position. This difference is due to the shift of the Fermi level of perfect borophene upward by up to 0.7 eV via electronic donation into boron according to our calculations. Taking such effect into account results in excellent agreement between the experiment and theory. The line defect is solely responsible for the appearance of this state, since the state will disappear once the line defect is removed in our calculations.

To explore the nature of this specific state, we plotted in Figure 3a the band structure of two (2,3) domains interconnected by two (2,2) ribbons. We identify two peculiar bands at ∼ 3.7 eV above the Fermi level that show parabolic dispersion relationship in the vicinity of the Γ point, colored in Figure 3a and denoted as the 1st and 2nd bands. The calculated effective masses ($m^*$) are 0.72 and 0.83 $m_e$ for these states, close to the free-electron mass ($m_e$), reminiscent of a NFE state. To confirm this point, we plot in Figure 3b its real-space charge distribution at Γ point. The partial charge density corresponding to the 1st band is found to hold to the material yet extend far into the vacuum region. In particular, the charge density is mainly distributed around the line defect and then extends evidently into the pure (2,3) phase, echoing the spatial variation of the state characterized experimentally. Moreover, the charge density of this state is several times sparser than other strongly bound states, such as the $π^*$ states (see the middle and bottom panel of Figure S2 in Supporting Information), further supporting the NFE nature. Based on this analysis, we attribute our experimentally observed delocalized state (~ -3.4 V) to the NFE state. We also identify several NFE states at higher energies with similar real-space distributions (an example for one of them is shown by green line in Figure S3, Supporting Information), which again agree with the fact that a series of Rydberg-like image potential states exist in layered nanomaterials.

After confirming that the delocalized electronic state around at the line defect is NFE state, we proceed by elucidating its physical origin. Similar to the NFE-like unoccupied image potential states on metal surfaces, the NFE state in our 1D domain boundary should extend in a 1D region parallel to line defect. As advanced by Chulkov and Silkin,[45] the image potential states are easy to be bound by Coulomb-like attractive potential well. Indeed, our calculations find an in-plane potential well centered at the line defect along the $x$ direction, as evidenced by the averaged potential shown in Figure 4. The potential was averaged in $y$-$z$ plane over the unit-cell area and plotted along the $x$ direction. This planar-averaged electrostatic potential exhibits periodic oscillation along the $x$ axis due to the spatial distribution of the electrons and ionic cores (shown as solid gray line in Figure 4). Whereas a macroscopic average of the potential avoids these oscillations and is shown by solid black or dashed red lines. A small slope of macroscopically averaged electrostatic potential is present at the abrupt interface. This discontinuity of the electrostatic potential produces a well, which can attract the weakly bound state (i.e. NFE state), manifested as downshift of their levels toward the Fermi level. Since the as-formed potential gradient only exist in the material plane, the NFE states are still adhered to the borophene plane and experience less perturbation along the normal of the plane. This forms a contrast to previously reported NFE states in other nanostructures, where the applied potential gradient (for example, by curvature, electric field or electron doping) is often perpendicular to materials basal planes.

With this mechanism, we conceive to deepen the potential well to shift the NFE state closer to the Fermi level, so that this state can take part in charge transport. Considering their spatial distribution, the electron transport via the NFE states can be directly modulated by variations in physical environments. Extensive calculations show that applying a 2% in-plane strain perpendicular to the line defect gives rise to a deeper potential well and thus make the NFE state located only slightly above the Fermi level.

**Conclusion.** In summary, we have realized the formation of a borophene sheet with line defects, formed by a structural unit of the (2,2) phase in the otherwise perfect (2,3) phase and observed for the first time highly delocalized, one-dimensional NFE states in the sheet using scanning tunneling microscopy/spectroscopy (STM/STS) characterizations. The characteristics and spatial distribution of the unoccupied NFE states were excellently reproduced by first principles calculations and the physical origin is elucidated. The rise of NFE state near the line defect in borophene is a unique feature apart from the metallic characteristic of borophene. Since the NFE states is unusually held to the atomic plane of borophene, and has a 1D feature, it may be applicable to tune the transport and electron emission properties of borophene-based devices.

**Figures**

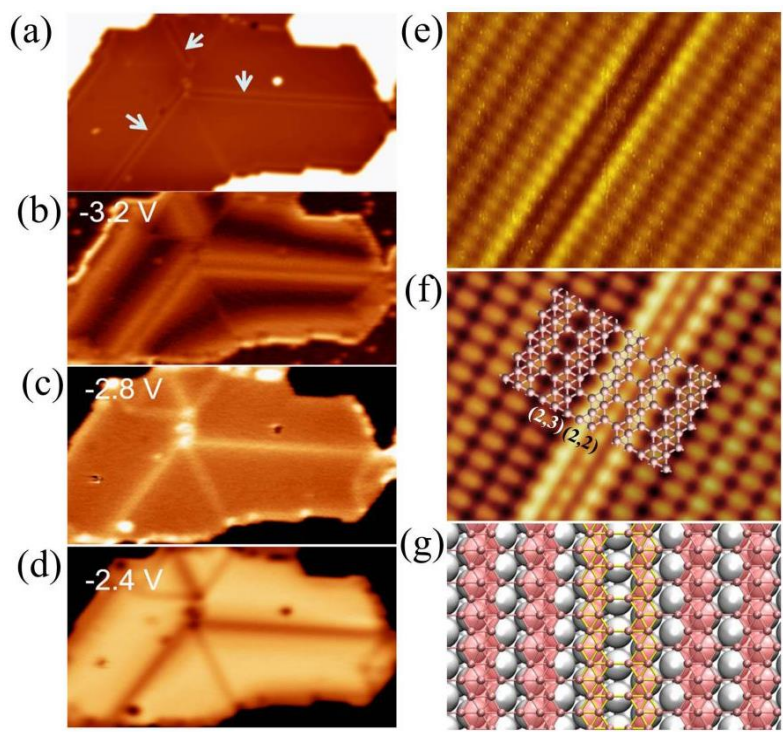

**Figure 1.** Topography of lateral borophene homojunctions on Ag(111). a) STM topographic image (25 nm × 50 nm) of (2,3) borophene phase with three linear defects (denoted by the white arrows) taken at bias -1.6 eV. b-d) The dI/dV maps taken at -3.2 V, -2.8 V and -2.4 V of the same borophene island as in (a). e) High resolution STM image (5.6 nm × 4.3 nm) of the surface including a typical line defect. f) Simulated STM image based on the structural model in (g), which is generated by using the calculated wave function from VASP in the energy range of 1.0 eV and visualized with an isovalue of 0.003 e/Å$^3$. g) DFT optimized structure model containing two (2,3) domains and separated by double (2,2) ribbons at the boundary. The (2,2) ribbons are highlighted by yellow stripes in g). The big and small balls represent silver and boron atoms respectively.

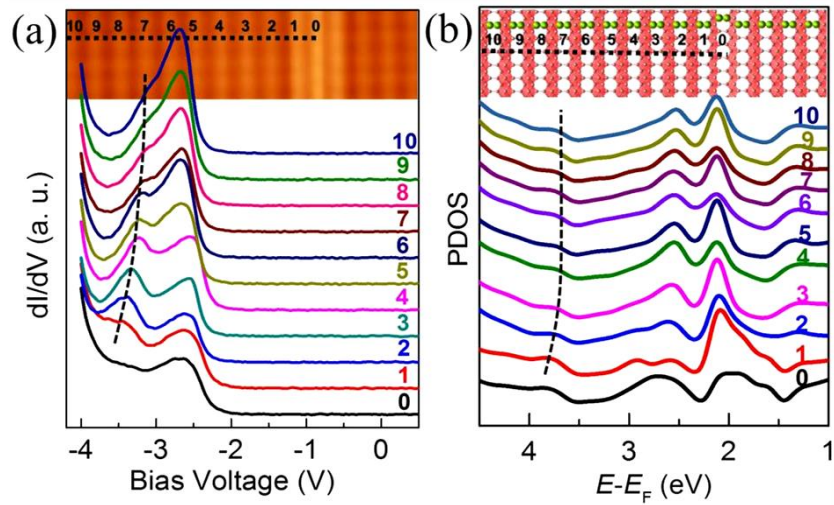

**Figure 2.** Line defect states of in-plane borophene homojunctions. a) Differential conductance spectra (dI/dV) taken at different boron chains marked by numbers in STM image (7.1 nm × 1.5 nm) shown at upper part. The chain with larger number is more far way to the line defects. b) The series of theoretical projected density of states (PDOS) obtained at different positions on borophene with double (2,2) ribbons line defects, which are marked by numbers. The positions with larger numbers are more far away to (2,2) ribbons. Protruding green balls in upper part indicates the projected atomic position of boron.

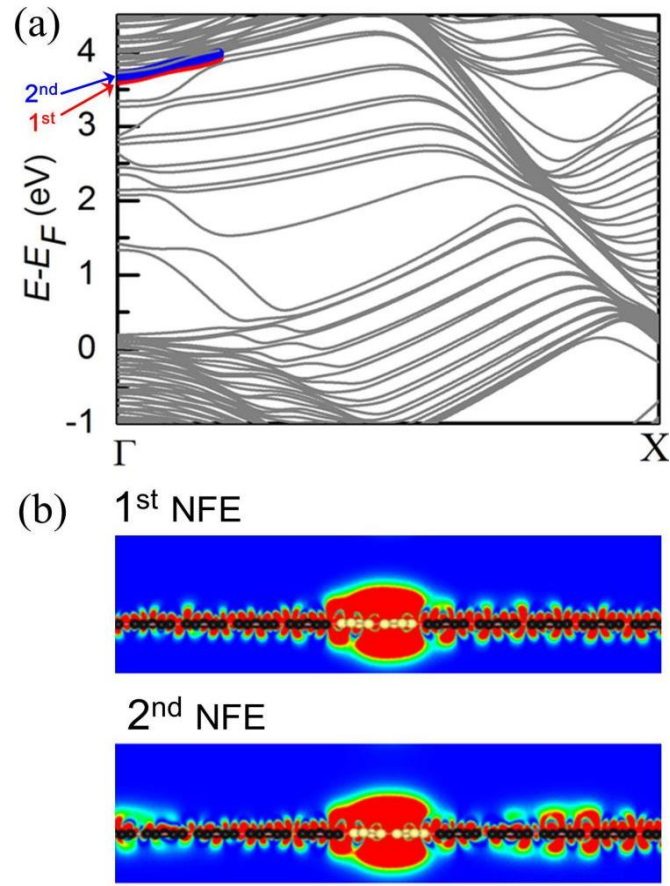

**Figure 3.** Effects of domain boundary on the electronic structures of lateral borophene homojunction from DFT calculations. a) Calculated band structure of freestanding borophene with linear defect. The energies are given with respect to the Fermi energy. The red, blue lines represent the 1st and 2nd NFE states, respectively. b) The side views (following the direction of the defect ribbons shown by yellow balls) of the real-space distribution of partial charge density at Γ point for the 1st and 2nd NFE states as marked by colored lines in (a). The (2,3) boron ribbons are indicated by black balls. The red color denote isosurface of charge density ($1\times10^{-6}$ e/Å$^3$).

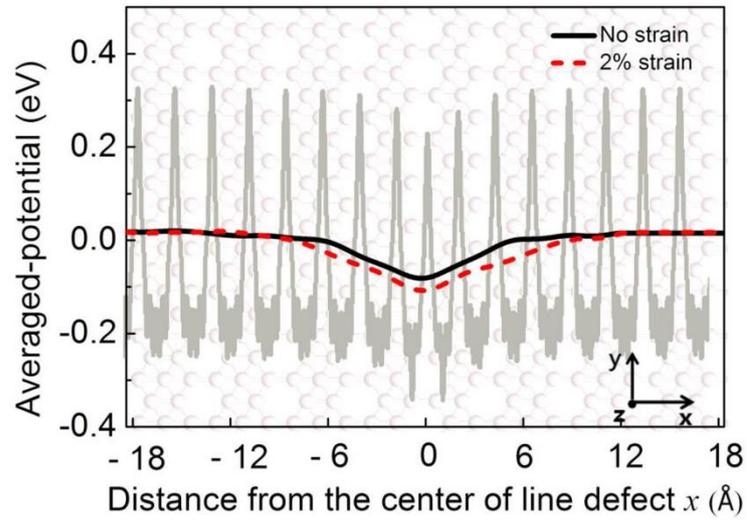

**Figure 4.** The planar-averaged electrostatic potential profile (gray periodic line) of the borophene film as a function of distance $x$ from the center of (2,2) ribbons as line defect. The solid black and dashed red lines represent the macroscopically averaged electrostatic potential under zero and 2% in-plane strain, respectively.

**Corresponding Author**

*E-mail: lchen@iphy.ac.cn (L. C.);

*E-mail: chuwazhang@nuaa.edu.cn (C. Z.);

*E-mail: khwu@iphy.ac.cn (K. W.)

**Author Contributions**

L. C. and K. W. designed the experiment; L. C. and Q. Z. grow the sample and preformed the STM experiments, as well as did the data analysis with K. W.; L. K., L. L., H. L. and Z. Z. designed and performed the first-principles calculations; L. K., L. C. and K. W. wrote the manuscript with contribution from all the authors; all co-authors contributed to analyzing and discussing the results.

**Funding Sources**

This work was supported by the MOST of China (grants nos. 2016YFA0300904, 2016YFA0202301), the NSF of China (grants nos. 11674366, 11761141013, 11825405, 11674368, 11772153, 21773005), the Beijing Natural Science Foundation (grants no. Z180007) and the Strategic Priority Research Program of the Chinese Academy of Sciences (grant nos. XDB30103000, XDB07010200).

**ACKNOWLEDGMENT**

H. L. thanks the supporting by the BUCT Fund for Disciplines Construction (XK1702). Z.Z. acknowledges the supporting by the Research Fund of SKL-MCMS (MCMS-0417G01), and a Project Funded by the Priority Academic Program Development of Jiangsu Higher Education Institutions.

**REFERENCES**

(1) Csányi, G.; Littlewood, P. B.; Nevidomskyy, A. H.; Pickard, C. J.; Simons, B. D. *Nat. Phys.* **2005**, 1, 42.

(2) Dougherty, D. B.; Feng, M.; Petek, H.; Yates Jr, J. T.; Zhao, J. *Phys. Rev. Lett.* **2012**, 109, 266802.

(3) Feng, M.; Zhao, J.; Petek, H. *Science* **2008**, 320, 359-362.

(4) Khazaei, M.; Ranjbar, A.; Ghorbani-Asl, M.; Arai, M.; Sasaki, T.; Liang, Y.; Yunoki, S. *Phys. Rev. B* **2016**, 93, 205125.

(5) Margine, E. R.; Crespi, V. H. *Phys. Rev. Lett.* **2006**, 96, 196803.

(6) Zhao, S.; Li, Z.; Yang, J. *J. Am. Chem. Soc.* **2014**, 136, 13313-13318.

(7) Yan, B.; Park, C.; Ihm, J.; Zhou, G.; Duan, W.; Park, N. *J. Am. Chem. Soc.* **2008**, 130, 17012-17015.

(8) Okada, S.; Oshiyama, A.; Saito, S. *Phys. Rev. B* **2000**, 62, 7634.

(9) Eknapakul, T.; Fongkaew, I.; Siriroj, S.; Vidyasagar, R.; Denlinger, J. D.; Bawden, L.; Mo, S. K.; King, P. D. C.; Takagi, H.; Limpijumnong, S.; Meevasana, W. *Phys. Rev. B* **2016**, 94, 201121(R).

(10) Feng, M.; Zhao, J.; Huang, T.; Zhu, X.; Petek, H. *Acc. Chem. Res.* **2011**, 44, 360-368.

(11) Otani, M.; Okada, S. *J. Phys. Soc. Jpn.* **2010**, 79, 073701.

(12) Posternak, M.; Baldereschi, A.; Freeman, A. J.; Wimmer, E. *Phys. Rev. Lett.* **1984**, 52, 863.

(13) Silkin, V. M.; Zhao, J.; Guinea, F.; Chulkov, E. V.; Echenique, P. M.; Petek, H. *Phys. Rev. B* **2009**, 80, 121408.

(14) Cuong, N. T.; Otani, M.; Okada, S. *J. Phys.: Condens. Matter* **2014**, 26, 135001.

(15) Yamanaka, A.; Okada, S. *Appl. Phys. Express*, **2014**, 7, 125103.

(16) Hu, S.; Zhao, J.; Jin, Y.; Yang, J.; Petek, H.; Hou, J. G. *Nano Lett.* **2010,** 10, 4830-4838.

(17) Zhao, J.; Feng, M.; Yang, J.; Petek, H. *ACS Nano* **2009,** 3, 853-864.

(18) Feng, B.; Zhang, J.; Zhong, Q.; Li, W.; Li, S.; Li, H.; Cheng, P.; Meng, S.; Chen, L.; Wu, K. H. *Nat. Chem.* **2016**, 8, 563.

(19) Mannix, A. J.; Zhou, X.-F.; Kiraly, B.; Wood, J. D.; Alducin, D.; Myers, B. D.; Liu, X.; Fisher, B. L.; Santiago, U.; Guest, J. R.; Yacaman, M. J.; Ponce, A.; Oganov, A. R.; Hersam, M. C.; Guisinger, N. P. *Science* **2015**, 350, 1513-1516.

(20) Mannix, A. J.; Kiraly, B.; Hersam, M. C.; Guisinger, N. P. *Nat. Rev. Chem.* **2017,** 1, 0014.

(21) Feng, B.; Zhang, J.; Ito, S.; Arita, M.; Cheng, C.; Chen, L.; Wu,; K.; Komori, F.; Sugino, O.; Miyamoto, K.; Okuda, T.; Meng, S.; Matsuda, I. *Adv. Mater.* **2018**, 30, 1704025.

(22) Li, W.; Kong, L.; Chen, C.; Gou, J.; Sheng, S.; Zhang, W.; Li, H.; Chen, L.; Cheng, P.; Wu, K. *Sci. Bull.* **2018**, 63, 282-286.

(23) Feng, B.; Sugino, O.; Liu, R.-Y.; Zhang, J.; Yukawa, R.; Kawamura, M.; Iimori, T.; Kim, H.; Hasegawa, Y.; Li, H.; Chen, L.; Wu, K.; Kumigashira, H.; Komori, F.; Chiang, T.-C.; Meng, S.; Matsuda, I. *Phys. Rev. Lett.* **2017**, 118, 096401.

(24) Kong, L.; Wu, K.; Chen, L. *Front. Phys.* **2018,** 13, 138105.

(25) Penev, E. S.; Kutana, A.; Yakobson, B. I. *Nano Lett.* **2016**, 16, 2522-2526.

(26) Huang, Y.; Shirodkar, S. N.; Yakobson, B. I. *J. Am. Chem. Soc.* **2017**, 139, 17181-17185.

(27) Zhang, J.; Zhang, J.; Zhou, L.; Cheng, C.; Lian, C.; Liu, J.; Tretiak, S.; Lischner, J.; Giustino F.; Meng, S. *Angew. Chem. Int. Ed.* **2018**, 57, 4585-4589.

(28) Zhang, Z.; Yang, Y.; Penev E. S.; Yakobson, B. I. *Adv. Funct. Mater.* **2017**, 27, 1605059.

(29) Mannix, A. J.; Zhang, Z.; Guisinger, N. P.; Yakobson B. I.; Hersam, M. C. *Nat. Nanotech.* **2018**, 13, 444-450.

(30) Zhang, Z.; Penev E. S.; Yakobson, B. I. *Chem. Soc. Rev.* **2017**, 46, 6746-6763.

(31) Piazza, Z. A.; Hu, H. S.; Li, W. L.; Zhao, Y. F.; Li, J.; Wang, L. S. *Nat. Commun.* **2014**, 5, 3113.

(32) Tang, H.; Ismail-Beigi, S. *Phys. Rev. Lett.* **2007**, 99, 115501.

(33) Penev, E. S.; Bhowmick, S.; Sadrzadeh, A.; Yakobson, B. I. *Nano Lett.* **2012**, 12, 2441-2445.

(34) Zhang, Z.; Yang, Y.; Guo, G.; Yakobson, B. I. *Angew. Chem. Int. Ed.* **2015**, 54, 13022-13026.

(35) Zhang, Z.; Penev, E. S.; Yakobson, B. I. *Nat. Chem.* **2016**, 8, 525.

(36) Wu, X.; Dai, J.; Zhao, Y.; Zhuo, Z.; Yang, J.; Zeng, X. C. *ACS Nano* **2012**, 6, 7443-7453.

(37) Zhong, Q.; Zhang, J.; Cheng, P.; Feng, B.; Li, W.; Sheng, S.; Li, H.; Meng, S.; Chen, L.; Wu, K. H. *J. Phys.: Condens. Matter* **2017**, 29, 095002.

(38) Zhong, Q.; Kong, L.; Gou, J.; Li, W.; Sheng, S.; Yang, S.; Cheng, P.; Li, H.; Wu, K.; Chen, L. *Phys. Rev. Mater.* **2017**, 1, 021001.

(39) Wu, R.; Drozdov, I. K.; Eltinge, S.; Zahl, P.; Ismail-Beigi, S.; Božović, I.; Gozar, A. *Nat. Nanotech.* **2019**, 14, 44.

(40) Liu, X.; Zhang, Z.; Wang, L.; Yakobson, B. I.; Hersam, M. C. *Nat. Mater.* **2018**, 17, 783-788.

(41) Perdew, J. P.; Burke, K.; Ernzerhof, M. *Phys. Rev. Lett.* **1996**, **77,** 3865.

(42) Kresse G.; Hafner, J. *Phys. Rev. B* **1993**, 47, 558.

(43) Kresse G.; Furthmüller, J. *Phys. Rev. B* **1996,** 54, 11169.

(44) Kresse G.; Joubert, D. *Phys. Rev. B* **1999,**59, 1758 .

(45) Chulkov, E. V.; Silkin, V. M.; Echenique, P. M. *Surf. Sci.* **1999**, 437, 330-352.